           \documentstyle[prl,preprint,aps]{revtex}% <--|  <----| <--|
\def\dm2{ \Delta m^2 }
\def\s22t{ \sin^22\theta }

\def\be{ \begin{equation} }
\def\ee{ \end{equation} }
\def\ba{\begin{array} }
\def\ea{\end{array} }
\def\bea*{ \begin{eqnarray*} }
\def\eea*{ \end{eqnarray*} }

\begin{document}

%latex%\renewcommand{\thefootnote}{\fnsymbol{footnote}}
%latex%\setcounter{footnote}{1}

%\draft

\title{
\hfill               {\normalsize \today
\\       \hfill                        UPR--0633T}
\\
       Four-Fermi Operators in $e^+$ $e^-$ Annihilation Experiments
       and Uncertainties in $Z$ Boson Properties
}
\author{
   Jens Erler
}

\address{
   Department of Physics,
   University of Pennsylvania, Philadelphia, PA 19104, U.S.A.
}

\maketitle

\begin{abstract}

We investigate the effects of possible new four-Fermi operators on
$e^+$ $e^-$ annihilation experiments. They represent a class of new physics,
which has the potential to change the parameters of the $Z$ boson and at the
same time interferes with their determination from the data. We show that in
the presence of such operators the $Z$ parameters obtained from lineshape
fits can change significantly. Another important property of these operators
is that they spoil the factorization of the expression for the left-right
asymmetry, $A_{LR}$, into initial and final state couplings. Factorization
and subsequent cancellation of the final state coupling occurs in the
Standard Model (after correcting for photonic amplitudes) and is crucial for
the interpretation of $A_{LR}$ as a measurement of the effective weak mixing
angle. Four-Fermi operators may thus provide an explanation for the high
value of the polarization asymmetry as observed at SLC. However, the data
from lower energy $e^+$ $e^-$ annihilation severely constrains this class of
operators and virtually closes this possible loophole on how new physics
might explain the SLC/LEP discrepancy. We point out that if the surplus of
observed $b$-pairs at LEP is real and caused by these operators, there may
be a significant effect on the forward-backward asymmetry into $b$-quarks.
It is this quantity which presently gives the most precise determination of
the weak mixing angle. We present compact analytical expressions for the
treatment of initial state radiation in the presence of new contact operators.

\end{abstract}
%\pacs{PACS numbers: 14.60.Gh, 96.60.Kx}

%\twocolumn
\newpage

%latex%\setlength{\textheight}{21.0cm}
%latex%\setlength{\topmargin}{-0.9cm}
%latex%\newpage
%latex%\vspace{4.0ex}

\section{INTRODUCTION:}

As the $Z$ factories LEP 1 and SLC increase their event samples and decrease
systematic uncertainties, many quantities related to the $Z$ boson are now
known with an impressive accuracy~\cite{Schaile}. A recent global
analysis~\cite{statussm} shows that the data are generally in impressive
agreement with the minimal Standard Model, though some observables show
deviations at the 2 or 3 $\sigma$ level and it cannot be excluded that
some of them are due to new physics.

At the heart of high precision experiments are the energy scans around
the $Z$ peak which are analyzed using various degrees of model independence.
One may work entirely in the context of the Standard Model, using as free
parameters the masses of the $Z$ and Higgs bosons and the top quark
in addition to the couplings $\alpha (M_Z)$ and $\alpha_s (M_Z)$.
Alternatively, one may allow for general vector and axial-vector couplings of
the $Z$ to fermions. Sometimes also more general interference terms are
admitted\footnote{See ref.~\cite{SMATASY} for a description of the new
ZFITTER~\cite{ZFITTER} interface SMATASY and review of other approaches.}.
All these approaches assume, however, that exclusively $Z$ and photon exchange
diagrams contribute to cross sections and asymmetries.

In this paper, we study the possibility that there may be additional
contributions to the cross sections (asymmetries) arising through new effective
four-Fermi operators. The $Z$ lineshape is known to be well described by
a Breit-Wigner curve which is only distorted by photonic contributions
and electroweak loop corrections with very little $s$-dependence. The $\chi^2$
values of the $Z$ lineshape fits are satisfactory and by themselves do not
call for the introduction of new parameters beyond some minimal set.
However, a major motivation for high precision experiments is the
hope to uncover new physics beyond the Standard Model. Likewise, one
wishes to extract limits on new physics, such as additional
$Z^\prime$s or compositeness. This is usually done by comparing the
observed $Z$ properties with the Standard Model expectations and with
expectations from certain kinds of new physics. But there is the logical
possibility that new interactions contribute to the event samples
at \mbox{$e^+$ $e^-$} annihilation experiments and obscure the
$Z$ studies. If new effects are present but not corrected for one
would extract erroneous parameters from the data.

Presently, the most important motivation for discussing new four-Fermi
interactions is the anomalously high left-right asymmetry, $A_{LR}$,
as observed by SLD~\cite{SLD2}.
We showed in ref.~\cite{Teupitz} that even the most general choice of
$Z$ couplings cannot significantly decrease the apparent discrepancy
between $A_{LR}$ and other high precision observables, and in particular
to the LEP asymmetries. Moreover, using results from reference~\cite{LLM} it
was argued~\cite{Teupitz} that no kind of new physics can account for
the SLD result without simultaneously conflicting with one or several other
observables, most notably the $W$ mass. Again, the one loophole in such a line
of argument could be a new interaction which significantly contributes
to the observed four fermion processes and so influences the extracted
$Z$ parameters. In the absence of new operators and after correcting for
photonic contributions, the asymmetries
factorize into expressions describing initial and final state couplings
(see Eqs.~(\ref{afb}) -- (\ref{ptaufb}) below). In the case of $A_{LR}$
the final state couplings then cancel and one is left with an expression
containing just the effective weak mixing angle for electrons.
On the other hand, in the presence of new operators factorization
and cancellation cease to hold so that the observables at LEP and SLC
describe inequivalent quantities.

At first glance it seems unlikely that there is any kind of new physics
which contributes significantly to $A_{LR}$ without simultaneously
affecting the high statistic LEP cross section measurements in an unacceptable
way. Surprisingly, as we discuss in section~\ref{4Fermiintfer}, $Z$
lineshape data alone cannot exclude a sizable contribution from new physics.
It is the data from lower energy $e^+$ $e^-$ annihilation experiments which
severely constrain this class of operators and virtually close a possible
loophole on how new physics might explain the SLC result.
There are no $Z$ interferences on the pole, but they arise away from
the pole and there they effect the factorization. Far away from the $Z$ pole
the interference of the new physics with the photon is in general
unsuppressed and this yields strong constraints for vector operators.

In section~\ref{4Fermigeneral} we collect some basic facts about effective
four-Fermi operators and discuss in detail (pseudo-) scalar and tensor
operators. Section~\ref{4Fermiintfer} focuses on (axial-) vector operators
and on how they might affect the lineshape measurements. For our analyses
we use the published cross section data from L3~\cite{L3}. In
section~\ref{conclusion} we summarize our conclusions. Appendix~\ref{approx}
describes the approximations we used for our fits. Explicit formulae for
initial state radiation in the presence of four-Fermi operators are presented
in appendix~\ref{ISR}.

\section{Four-Fermi Operators and Helicity Amplitudes:}
\label{4Fermigeneral}

The most general four-fermion contact operator has the form~\cite{Scheck}
\be
\label{lint}
   {\cal L}_{\rm eff} = - {4\pi \over \Lambda^2} \sum\limits_{i=S,P,V,A,T}
   \bar{f}_1 \Gamma^i f_2 (D_i^{f_1,f_2,f_3,f_4} \bar{f}_3 \Gamma_i f_4 +
   \tilde{D}_i^{f_1,f_2,f_3,f_4} \bar{f}_3 \Gamma_i \gamma_5 f_4) + h.\, c.,
\ee
where the sum is over scalar, pseudo-scalar, vector, axial-vector and
tensor operators. The scale $\Lambda$ is introduced for convenience and
later it will be taken to be 1 TeV. For $e^+$ $e^-$ annihilation we set
$f_1 = f_2 = e^-$. For definiteness we restrict ourselves to
flavor conserving neutral currents\footnote{We expect our conclusions
to hold for the case $f_3 \neq f_4$, as well.} so that $f_3 = f_4$.
Then the effective Lagrangian (\ref{lint}) can be rewritten as
\be
\label{lint2}
   {\cal L}_{\rm eff} = - {4\pi \over \Lambda^2} \sum\limits_{i,j = L,R}
   (S_{ij}^f \bar{e}^- P_i e^- \bar{f} P_j f +
    V_{ij}^f \bar{e}^- \gamma^\mu P_i e^- \bar{f} \gamma_\mu P_j f +
    T_{ij}^f \bar{e}^- {\sigma^{\mu\nu} \over \sqrt{2}} P_i e^-
            \bar{f} {\sigma_{\mu\nu} \over \sqrt{2}} P_j f) + h.\, c.,
\ee
where $P_i = (1 + h_i \gamma_5)/2$ with $h_i = \mp 1$ for left and right-handed
fermions, respectively. The coefficients $S_{ij}^f$ and
$T_{ij}^f$ satisfy the relations
\be
\label{coeff}
\ba{ll}
   S_{LL}^f = {S_{RR}^f}^*, \\
   S_{LR}^f = {S_{RL}^f}^*, \\
   T_{LL}^f = {T_{RR}^f}^*, \\
   T_{LR}^f = T_{RL}^f  = 0,
\ea
\ee
whereas the $V_{ij}^f$ are real\footnote{On the pole interferences
with the $Z$ are only possible for complex couplings, but these are not
allowed for $f_3 = f_4$. If $f_3 \neq f_4$ there is no interference due to the
absence of flavor changing neutral currents in the Standard Model.}.

The helicity amplitudes for $e^+$ $e^-$ annihilation
through (axial-) vector channels are given by
\be
\ba{rcl}
   \frac{{\rm d} \sigma_{ijf}^V}{{\rm d} z}(s) &=& {\pi N_c s\over 8}
   |\alpha \frac{g_i^e g_j^f}{s-M_Z^2+i M_Z \Gamma_Z}
            + \alpha \frac{Q_e Q_f}{s} + \frac{V_{ij}^f}{\Lambda^2}|^2
   (1 + h_i h_j z)^2 \\
    &\equiv& {\pi N_c s\over 8} |A_{ij}^f|^2 (1 + h_i h_j z)^2,
\ea
\ee
where $z=\cos \theta$. The first term of the amplitude is the $Z$ contribution,
where $g_i^f$ describes the effective coupling of the $Z$ to fermion $f$ with
helicity $h_i$. The second term is the contribution from QED and the third
one from new physics.

If we now define cross sections for a given fermion flavor
of specific helicity to be detected in the forward or backward hemisphere,
\be
\ba{l}
  {\sigma_{ijf}^V}^F \sim \int\limits_{ 0}^{1} {\rm d}z [1 + h_i h_j z]^2, \\
  {\sigma_{ijf}^V}^B \sim \int\limits_{-1}^{0} {\rm d}z [1 + h_i h_j z]^2,
\ea
\ee
we find for forward-backward asymmetries
\be
   A_{FB} (i,j,f) \equiv \frac{\sigma_{ijf}^F - \sigma_{ijf}^B}
        {\sigma_{ijf}^F + \sigma_{ijf}^B} = {3 \over 4} h_i h_j.
\ee
Averaging over initial and summing over final helicity states yields
\be
\label{afb}
   A_{FB} (f) = {3 \over 4}
   \frac{|A_{LL}^f|^2 + |A_{RR}^f|^2 - |A_{LR}^f|^2 - |A_{RL}^f|^2}
   {|A_{LL}^f|^2 + |A_{RR}^f|^2 + |A_{LR}^f|^2 + |A_{RL}^f|^2} = {3 \over 4}
   \frac{({g^e_L}^2-{g_R^e}^2)({g_L^f}^2-{g_R^f}^2)}
    {({g_L^e}^2 + {g_R^e}^2)({g_L^f}^2 + {g_R^f}^2)}
   \equiv {3 \over 4} A_e A_f,
\ee
where the second equality holds when one ignores any photonic or new physics
contributions, in which case the result is independent of $s$.
All $s$-dependence enters through the photon
amplitude and possibly through new interactions.

Similarly, one has for the left-right asymmetry
\be
\label{alr}
   A_{LR} (f) =
   \frac{|A_{LL}^f|^2 + |A_{LR}^f|^2 - |A_{RL}^f|^2 - |A_{RR}^f|^2}
        {|A_{LL}^f|^2 + |A_{LR}^f|^2 + |A_{RL}^f|^2 + |A_{RR}^f|^2} =
%   \frac{({\bar{g}_L^e}^2-{\bar{g}_R^e}^2)({\bar{g}_L^f}^2+{\bar{g}_R^f}^2)}
%   {({\bar{g}_L^e}^2+{\bar{g}_R^e}^2)({\bar{g}_L^f}^2+{\bar{g}_R^f}^2)} = A_e.
   \frac{({g_L^e}^2-{g_R^e}^2)({g_L^f}^2+{g_R^f}^2)}
   {({g_L^e}^2+{g_R^e}^2)({g_L^f}^2+{g_R^f}^2)} = A_e,
\ee
and it becomes obvious that the final state couplings drop out even after
summation over final state fermion flavors\footnote{SLD counts hadronic and
$\tau$-events.}. After inclusion of photonic or new physics amplitudes the
factorization and cancellation of the final state couplings ceases to hold.

The $\tau$-polarization and its forward-backward asymmetry
are given by
\be
   {\cal P}_\tau =
   \frac{|A_{RR}^\tau|^2+|A_{LR}^\tau|^2-|A_{RL}^\tau|^2-|A_{LL}^\tau|^2}
        {|A_{RR}^\tau|^2+|A_{LR}^\tau|^2+|A_{RL}^\tau|^2+|A_{LL}^\tau|^2} =
   \frac{{g_R^\tau}^2-{g_L^\tau}^2}
        {{g_R^\tau}^2+{g_L^\tau}^2} = - A_\tau,
\ee
and
\be
\label{ptaufb}
   {\cal P}_\tau^{FB} = {3 \over 4}
   \frac{|A_{RR}^\tau|^2-|A_{LR}^\tau|^2+|A_{RL}^\tau|^2-|A_{LL}^\tau|^2}
        {|A_{RR}^\tau|^2+|A_{LR}^\tau|^2+|A_{RL}^\tau|^2+|A_{LL}^\tau|^2} =
%   - {3 \over 4} \frac{{\bar{g}_R^e}^2-{\bar{g}_L^e}^2}
%        {{\bar{g}_R^e}^2+{\bar{g}_L^e}^2} = - {3 \over 4} A_e,
   - {3 \over 4} \frac{{g_R^e}^2-{g_L^e}^2}
                      {{g_R^e}^2+{g_L^e}^2} = - {3 \over 4} A_e,
\ee
respectively. Thus, after correcting for all photonic effects (including
initial state radiation) $A_{LR} (\tau)$ and ${\cal P}_\tau^{FB}$ are
physically equivalent quantities. Moreover, even in the presence of photonic
and/or new physics contributions to the amplitude they are identical
functions\footnote{They receive slightly different corrections from
initial state radiation, though.}
of $s$. In practice, however, less than 1\% of the SLD sample consists of
$\tau$-pairs~\cite{SLD2}.

Let us compare this with the case of (pseudo-) scalar interactions for which
we find
\be
   \frac{{\rm d} \sigma_{ijf}^S}{{\rm d} z}(s) = {\pi N_c s\over 8}
   |\frac{S_{ij}^f}{\Lambda^2}|^2.
\ee
%where
%\be
%   A_{ijf}^S = \frac{S_{ij}^f}{\Lambda^2}.
%\ee
Notice, that this result is independent of the scattering angle $\theta$
as well as the helicity structure. The first property trivially shows that
all forward-backward asymmetries vanish for this class of operators.
The second property combined with the hermiticity conditions (\ref{coeff})
for the $S_{ij}^f$ shows that the left-right asymmetries vanish as
well\footnote{The only non-trivial asymmetry would be a
combined initial and final state polarization asymmetry, i.e., a measurement
of final state helicities using a polarized beam. This kind of asymmetry has
not been measured, yet.}. On the other hand, such operators contribute to total
cross sections and consequently they can lower the asymmetries from
(axial-) vector type interactions.

Is it possible that LEP asymmetries are lowered more significantly
in this way than the SLD asymmetries?
Presently, the most precise determination of the weak mixing angle comes from
$A_{FB} (b)$, which corresponds to
\be
  A_e [A_{FB} (b)] = 0.1380 \pm 0.0051,
\ee
and which is to be compared with
\be
  A_e [A_{LR}] = 0.1637 \pm 0.0075.
\ee
Thus the two most precise determinations show the largest discrepancy
(2.8 $\sigma$). So let us assume that there is an additional contribution
to the cross section into $b$-quarks effectively lowering
$A_{FB} (b)$ such as to resolve the discrepancy.
Neglecting the $\tau$-pairs at SLC we find for
the fraction of $b$-events due to new physics
\be
   \frac{\sigma_b^{\rm new}}{\sigma_b} = \frac{A_e [A_{LR}]-A_e [A_{FB} (b)]}
                    {A_e [A_{LR}]-R_b A_e [A_{FB} (b)]} \approx 0.19,
\ee
where we used the LEP value $R_b = \sigma_b/\sigma_{had} = 0.2192 \pm 0.0018$
for $R_c$ fixed to its Standard Model value. This would correspond to
\be
\label{aeconv}
   A_e^{\rm SM}=\frac{A_e [A_{FB} (b)]}{1-\frac{\sigma_b^{\rm new}}{\sigma_b}}
       = \frac{A_e [A_{LR}]}{1-R_b \frac{\sigma_b^{\rm new}}{\sigma_b}}
         \approx 0.171,
\ee
and to $m_t \approx 270$ GeV (for $M_H = 300$ GeV).
We use this value of $m_t$ to compute
\be
\label{rbconv}
   R_b = \frac{R_b^{\rm SM}}
      {1+(R_b^{\rm SM}-1)\frac{\sigma_b^{\rm new}}{\sigma_b}} \approx 0.249,
\ee
which is in clear conflict with the value from LEP. In any case, such a high
value of $m_t$ would also be in sharp conflict with other observables, such as
the $W$ mass and, of course, with the top-quark interpretation of the
CDF events.

In conclusion, (pseudo-) scalar four-Fermi operators cannot resolve the
LEP/SLC discrepancy and have little impact on the asymmetries.
Still, it is interesting to note that in the presence of
new physics which produces additional $b$-quarks (1) the LEP/SLC
discrepancy becomes slightly smaller, (2) the $R_b$ measurement is accounted
for, and (3) the high values for $\alpha_s$ from $R_{had}$ are
lowered~\cite{statussm} to be in better agreement with most low energy
determinations.

The next class of effective operators to be discussed are of (pseudo-) tensor
type. In this case we find for the cross sections,
\be
   \frac{{\rm d} \sigma_{ijf}^T}{{\rm d} z}(s) = {\pi N_c s\over 4}
   |\frac{T_{ij}^f}{\Lambda^2}|^2 (1 + h_i h_j) z^2.
\ee
%with
%\be
%   A_{ijf}^T = \frac{T_{ij}^f}{\Lambda^2}.
%\ee
Since the cross sections are proportional to $\cos^2 \theta$ there is
again no forward-backward asymmetry for this class of operators.
In accordance with the last relation (\ref{coeff}) we see that
helicity changing amplitudes vanish. The hermiticity condition on
the coefficients for helicity conserving processes implies the
vanishing of $A_{LR}$, as well.
Thus the same conclusions apply for (pseudo-) tensor operators as for the
(pseudo-) scalar class.

The situation changes when both
types of operators are present simultaneously. In this case interference terms
are possible for helicity conserving processes,
\be
   \frac{{\rm d} \sigma_{ijf}^{S+T,I}}{{\rm d} z}(s) =
   -{\pi N_c s\over 4\Lambda^4} {\rm Re}\; (S_{ij}^f {T_{ij}^f}^*)(1+h_i h_j)z.
\ee
Since these terms are linear in $\cos\theta$, we find non-trivial contributions
to forward-backward asymmetries if the coefficients are not out of phase.
The cross section for combined scalar and tensor interactions is
\be
   \frac{{\rm d} \sigma_{ijf}^{S+T}}{{\rm d} z}(s) =
   {\pi N_c s\over 8\Lambda^4} |S_{ij}^f - T_{ij}^f (1+h_i h_j) z|^2.
\ee
{}From this expression it can again be shown that $A_{LR}$ vanishes
identically. The (unnormalized) left-right asymmetry, $\sigma_L - \sigma_R$,
is sensitive precisely to the (axial-) vector part of the theory.
The forward-backward asymmetry, on the other hand, takes on the form
\be
   A_{FB}^{S+T} = \frac{- 2 {\rm Re}\; (S_{LL}^f {T_{LL}^f}^*)}
                 {|S_{LL}^f|^2 + |S_{LR}^f|^2 + {4\over 3} |T_{LL}^f|^2}.
\ee
It takes its maximal negative value for $S_{LR}^f = 0$ and $T_{LL}^f
= {\sqrt{3}\over 2} S_{LL}^f$, in which case it is
\be
  A_{FB}^{S+T,{\rm max}} = - {\sqrt{3}\over 2} \approx -86.6\% .
\ee
It should be noted that the new physics contributions considered here always
spoil the factorization properties (second equal signs in Eqs.~(\ref{afb})
-- (\ref{ptaufb})) trivially in that they contribute to the total cross
section in the denominators. But here the factorization is affected in addition
through a direct contribution to $\sigma^F - \sigma^B$.

We have to ask again, whether it is possible that LEP asymmetries are changed
more significantly than $A_{LR}$. Suppose that
$\frac{\sigma_b^{\rm new}}{\sigma_b} \approx 3.3\%$. Then using
Eq.~(\ref{aeconv}) yields $A_e^{\rm SM} \approx 0.165$. This corresponds
to $m_t = 254$ GeV (again for $M_H =  300$ GeV) and in this case
Eq.~(\ref{rbconv}) is
satisfied for the LEP value of $R_b$. The forward-backward asymmetry
into $b$-quarks would now be given by the cross section weighted average
\be
   A_{FB}(b)= (1 - \frac{\sigma_b^{\rm new}}{\sigma_b}) A_{FB}^{SM}(b)
                 + \frac{\sigma_b^{\rm new}}{\sigma_b}  A_{FB}^{S+T}(b).
\ee
Using $A_{FB}(b)=0.0967 \pm 0.0038$ from LEP and $A_{FB}^{SM}(b) = 0.1157$
we find $A_{FB}^{S+T}(b) \approx -46\%$. This is clearly acceptable, but there
remains the conflict between $A_{LR}$ and $A_e$ from ${\cal P}_\tau^{FB}$
which would now be about 2.3 $\sigma$. The point is that final state
asymmetries are as insensitive to these operators as initial state asymmetries.
In order to solve this conflict,
one would have to assume that about 18\% of the cross section into $\tau$-pairs
is due to new physics. This would be in sharp conflict with the
$Z \rightarrow \tau^+ \tau^-$ width, which is determined to be in excellent
agreement with the Standard Model value. Including  $A_\tau$ from
${\cal P}_\tau$ increases this discrepancy to 2.5 $\sigma$; still an
admixture of 16\% $\tau$-pairs from new physics would be required.
Even if one assumes that the $\tau$-polarization measurements are principally
wrong and dismisses them, one still faces the above mentioned conflicts
between a high $m_t$ on one hand and $M_W$ and CDF on the other.

In summary, we have shown that the data do not support the hypothesis
that (pseudo) scalar and tensor four-Fermi operators may resolve the
discrepancy of $A_{LR}$ with other asymmetries. These operators do not
contribute to polarization asymmetries and can at best resolve conflicts
between forward-backward asymmetries and polarization asymmetries, but not
between different polarization asymmetries. Moreover, as shown
they can only lower the measured asymmetries and hence increase the
extracted bare asymmetries. This would lead to very high values of $m_t$ which
are inconsistent with many other observables. On the other hand, if the
excess of $b$-quarks over the Standard Model value as measured at LEP
is real and due to these operators, the extracted value of
$\sin^2 \theta_{\rm eff}^e$ from $A_{FB}(b)$ could be significantly
lower or higher. This is particularly interesting in view of the fact that
by now $A_{FB}(b)$ yields the most precise determination of the weak mixing
angle.

\section{VECTOR AND AXIAL-VECTOR OPERATORS:}
\label{4Fermiintfer}

In the previous section we discussed the four helicity amplitudes related
to (pseudo-) scalar and tensor operators. They are the ones describing
interactions between states of opposite helicities. There are four other
amplitudes for interactions between states of equal helicities.
There cannot be interferences between the eight different helicity structures.
The new features related to (axial-) vector operators are that they can
interfere with the photon and the $Z$ and that they lead to non-trivial
polarization asymmetries. Also, they are theoretically better motivated.
E.g., they may arise from heavy $Z^\prime$ bosons, whose presence would not
spoil the successful supersymmetric gauge coupling unification. In that
case new effects can show up in two different ways: (1) $Z$ -- $Z^\prime$
mixing can change the mass and coupling relations of the ordinary $Z$.
(2) The determination of $Z$ properties from the data is modified.

It is often believed that such new operators cannot significantly
contribute to the high statistics and high precision measurements at the
$Z$ pole since the $Z$ lineshape is measured so well.
But the approximate Breit-Wigner curve only proves that
the $Z$ dominates the measurements. Other contributions could still
affect the details of the fit (parameters). It is also not appropriate to
conclude from the decent $\chi^2$ values from the lineshape fits that there
cannot be significant contributions from new sources. It is amusing to note
that for the L3 cross section data $\chi^2$ actually decreases by 0.8 when
photon exchange and interference are omitted\footnote{In order to avoid
complications with $t$-channel exchange we omit the data for final state
electrons.}! In such a fit, the $Z$ width and the leptonic partial width
increase by about 10 and 1 MeV, respectively, whereas the hadronic bare cross
section, $\sigma^0_{\rm had}$, and $M_Z$ remain unchanged.
This exercise should warn us that new physics which couples as strong as the
photon or even stronger could have been overlooked and significantly influence
lineshape parameters.

Suppose now that there are new purely vector-like four-Fermi operators
for hadrons only ($V_{ij}^q = V^q \neq 0$, $V_{ij}^l = 0$).
This is a particularly interesting case for the following reasons:
\begin{enumerate}
\item
    The anomalously high 1993 data at SLC was taken at $\sqrt{s} = 91.26$ GeV,
    whereas the low statistics run of 1992 at $\sqrt{s}=91.55$ GeV~\cite{SLD1}
    yielded a 1.5 $\sigma$ lower result for $A_{LR}$. At first sight this does
    not seem to be a possible hint that we see some (additional)
    $s$-dependence from interference terms. It seems that the critical 1993
    measurement was closer to $M_Z$ than the 1992 run and hence that new
    interferences would be more strongly suppressed in 1993. However,
    initial state radiation effects effectively shift the peak position by
    more than 200 MeV to higher energies, so that interference effects in 1993
    are larger and in 1992 smaller than naively expected.
\item
    The interference terms with the $Z$ contributing to $A_{LR}$ are
    proportional to the axial-vector coupling, $g_A^e$, whereas the ones
    contributing to total
    cross sections are suppressed by the small vector coupling, $g_V^e$.
\item
    We have to require destructive interference of the new physics with the
    photon in order to avoid too large contributions to the hadronic cross
    section;
    this fixes the sign of $V^q$ to be the same as the sign of $Q_q$.
    With this choice the $Z$ interference effect which might explain the
    high $A_{LR}$ also has the right sign.
\item
    We included these operators into a $Z$ lineshape fit using L3 cross
    section data and assuming family universality ($V^u = V^c$ and $V^d = V^s =
    V^b$) we found a $\chi^2$ minimum for $V^u \sim V^d \sim 0.5$, although
    the result was consistent with $V^q = 0$. Note, that the values
    $V^u = 0.585$ or $V^d = 0.293$ correspond to an interaction strength
    comparable to QED.
\item
    The combined fit of the L3 and SLC data gives a significant $\chi^2$
    minimum at $V^u = 1.2 \pm 1.0$ and $V^d = -1.2 \pm 0.6$. Requiring
    a maximal value for $A_{LR}$ gives virtually the same answer, showing
    that cross section measurements around the $Z$ pole are not very sensitive
    to these new operators and that the fit result is dominated by the SLD
    asymmetry data.
\end{enumerate}
The maximal effect for $A_{LR}$ comes about because for higher
values of the $V^q$ the contribution to the total hadronic cross
section would overcompensate the interference effect. We found shifts
of $\pm 8$ MeV in $M_Z$ and $\Gamma_Z$, respectively, whereas
$\sigma^0_{\rm had}$ and $\Gamma_l$ decreased by less than a standard
deviation.

Allowing these new operators increases the theory value of $A_{LR}$ (91.26 GeV)
from 0.1385 to 0.1469,
which is still low compared to the measured value of 0.1656.
About a third of the discrepancy could be accounted for, but it is still
2.1 $\sigma$. However, it is the older data from $e^+$ $e^-$
annihilation experiments well below the $Z$ which eliminates
this scenario. At the $Z$ peak the new physics would be stronger than
QED. Because of the destructive interference with the photon
one would predict that at some
energy below the $Z$ the combined amplitude from QED and the new physics
would vanish. At $\sqrt{s} = 34$ GeV, QED would be stronger
than the new physics but still one would expect only about half of the
observed hadronic cross section.
What has originally been seen was actually a slight enhancement of the cross
section over the Standard Model prediction, but a new analysis by
Haidt~\cite{Haidt} now shows agreement.

Henceforth, we will assume
that new physics interference with the photon vanishes, i.e., we require
\be
\label{nointerf}
   V_{LL}^f + V_{LR}^f + V_{RL}^f + V_{RR}^f = 0.
\ee

In the remainder of this section we have a closer look at two representative
cases. The first case (a) focusses on the possibility that $A_{LR}$ may be
enhanced by interference effects, like for the pure vector operators discussed
before. The second case (b) assumes that the new
physics couples to left-handed electrons only, which obviously would enhance
$A_{LR}$, as well.

Since case (a) concentrates on interference terms to $A_{LR}$ we choose
$V_{LL}^f = V_{RL}^f$ in order to retain the enhancement of
$g_A^e/g_V^e$ compared to the cross section interference terms. In order
to satisfy condition~(\ref{nointerf}) we further choose
\be
   V_{LR}^f = V_{RR}^f = - V_{LL}^f = -V_{RL}^f.
\ee
Finally, we assume family universality and
in view of the conflict between $A_{LR}$ and ${\cal P}_\tau^{FB}$
we restrict the new physics to contribute to hadrons only,
\be
\label{taucond}
   V_{ij}^l = 0.
\ee
Using exclusively L3 data actually yields a lower value of $\chi^2$ by about
0.7. The fit result on the two new parameters, however, are such that
$A_{LR}$ would actually be decreased. Including the SLC data then gives
values for the new parameters very close to zero and hence we dismiss case (a)
as uninteresting.

To maximize positive contributions to $A_{LR}$ in case (b) we set
\be
   V_{RL}^f = V_{RR}^f = 0.
\ee
Interestingly, this suppresses contributions to
forward-backward asymmetries and final state polarizations.
Again we assume that condition~(\ref{taucond}) holds.
Finally, motivated by the large fraction of $b$-events observed at LEP we set
\be
\ba{l}
   V_{LL}^b = -V_{LR}^b \neq 0, \\
   V_{ij}^f = 0 \;\;\; {\rm otherwise}.
\ea
\ee
There are interferences in case (b), as well, but their $g_A^e/g_V^e$
enhancement for $A_{LR}$ no longer holds.

Using the LEP values $\sin^2\theta_{\rm eff}^e = 0.2321 \pm 0.0004$
and $\frac{\sigma_b^{\rm new}}{\sigma_b} \approx 2\%$ and again
neglecting the $\tau$-events at SLC we find
\be
   A_{LR}^0= (1 - \frac{\sigma_b^{\rm new}}{\sigma_b} R_b) A_{LR}^{SM}
                + \frac{\sigma_b^{\rm new}}{\sigma_b} R_b = 0.1462,
\ee
which is still far lower than the measured value. In fact, we would need
a fraction $\frac{\sigma_{\rm had}^{\rm new}}{\sigma_{\rm had}}
\approx 2.5\%$ of hadrons from new physics to account for the SLD result.
Given the fact that $\sigma_{\rm had}^0 = 41.49 \pm 0.12$ [nb] and
$R_l = \frac{\sigma_{\rm had}^0}{\sigma_l^0} = 20.795 \pm 0.040$
are measured with 3 and 2 per mill accuracy, respectively, does not leave
much room for achieving that. But as mentioned before, a new $Z$ lineshape
fit including the new operators might change the picture:

We allow the new physics to couple to all quark flavors in a family
universal way ($V_{LL}^d = -V_{LR}^d = V^d$ and likewise for $V^u$).
Using only L3 data we find $V^u = - 0.9 \pm 1.4$ and $V^d = -0.5 \pm 0.9$
with a slight decrease in $\chi^2$ ($-0.3$ compared to the Standard Model fit).
Including the SLC data yields $V^u = -1.9 \pm 0.6$ and $V^d = -1.1 \pm 0.3$.
Note, that the obtained values are consistent with the $SU(2)$ symmetric case,
$V^u = V^d$. In such a fit we find for the lineshape parameters,
\be
\ba{llrcl}
\label{fitparam}
   M_Z &=& 91.184 &\pm& 0.012\; {\rm GeV}, \\
   \Gamma_Z &=& 2.468 &\pm& 0.020\; {\rm GeV}, \\
   \sigma_{\rm had}^0 &=& 41.19 &\pm& 0.25\; {\rm nb}, \\
   \Gamma_l &=& 82.74 &\pm& 0.64\; {\rm MeV}.
\ea
\ee
They deviate significantly from their Standard Model counterparts. In this case
only about 21\% of the $A_{LR}$ discrepancy can be accounted for and still
amounts to 2.5 $\sigma$. The discrepancy is actually a little smaller than that
because of the effect of the additional $b$-quarks on $A_{FB}(b)$ as discussed
in the previous section. Although this scenario is interesting in that it
describes the data better than the Standard Model, we find only a modest
increase in $A_{LR}$. Actually, it turned out that these operators
may rather account for the large observed partial $Z$ width into
$b$-quarks. The value of $V^d = -1.1$ corresponds to a fraction of
about 1\% $b$-quarks due to new physics. Cross sections into up-type quarks
are rather lowered (which would also be consistent with observation),
but not significantly so. These new operators may have an effect on low energy
data as well. But since by demand they do not interfere with QED
this effect is much smaller than in the pure vector case, and there would be
an enhancement in the hadronic cross section. We do not use these low energy
data to quantitatively constrain these operators, since there may be
competing residual interferences or the new physics may actually decrease
faster towards lower energies than is the case for a pure four-Fermi
operator.

\section{Conclusions}
\label{conclusion}

We can finally conclude that $e^+$ $e^-$ data do not support the idea that
the presence of new four-Fermi operators might solve the conflict between
SLC and LEP asymmetry measurements. (Pseudo-) scalar and tensor operators
do not give rise to polarization asymmetries. Forward-backward asymmetries
are only possible when helicity conserving scalar and tensor interactions
interfere. If the observed surplus of $Z \rightarrow b\bar{b}$ events is
real and due to this class of operators the value of $\sin^2\theta_{\rm eff}^e$
extracted from $A_{FB}(b)$ may significantly change. This is important since
presently $A_{FB}(b)$ serves as the most precise determination of the weak
mixing angle.

(Axial-) vector operators are constrained to have basically
no interference with the photon. We obtained this constrained by
considering lower energy $e^+$ $e^-$ annihilation data and using it
we showed that
the interference effect with the $Z$ cannot explain the large left-right
asymmetry. If the new physics couples predominantly to left-handed
electrons there is a sizable effect on $A_{LR}$, but less than a fourth of
the discrepancy can be accounted for. Still this class of operators
is interesting since (1) it may account for the surplus of observed
$b\bar{b}$ final states; (2) it has significant impact on extracted $Z$ pole
parameters, in particular total and partial widths; and (3) it can account
at least for part of the $A_{LR}$ puzzle.

New (axial-) vector operators could for instance arise through a
new $Z^\prime$-boson. This case is discussed in a very recent paper by
Caravaglios and Ross~\cite{CR}. They conclude that only an (almost) degenerate
$Z^\prime$ can affect the SLAC measurement while leaving the LEP observables
unaffected. Indeed, a $Z^\prime$ resonating close to the $Z$ is only
poorly described by four-Fermi interactions since its amplitude
falls off much faster at lower energies than is the case for
contact operators. Moreover, it strongly interferes {\em at\/} the pole
and not only {\em near\/} it. On the other hand, it should be possible
to further constrain the strength of its couplings by considering its
interference with the photon at lower energies. Although the pure photon
exchange dominates at energy scales such as at PETRA, the $Z$ already
contributes significantly, mainly through its interference with QED.
The size of the analogous effect of $Z^\prime$ -- $\gamma$ mixing
for a $Z^\prime$ such as considered in~\cite{CR} depends on how it
couples to fermions. There is, however, a relative enhancement of this effect
as the $Z^\prime$ is described as predominantly vector-like. Thus it would be
interesting to study its implications at lower energies.

In view of fit results such as~(\ref{fitparam}) we must further conclude that
in the presence of certain kinds of new physics our knowledge of $Z$ boson
properties may be poorer than expected. The generally low $\chi^2$ values of
$Z$ lineshape fits should not be taken as a guarantee that new physics
contributions are necessarily negligible.
Therefore we encourage the LEP collaborations to perform in addition to their
usual ``model independent fits'' more general fits allowing new physics such
as the kind discussed in this paper, in which we could only attempt
a semi-quantitative discussion. We were forced to use a number of
approximate treatments for our fits as described in appendix~\ref{approx}.
Despite of all these simplifications,
our agreement with L3 is very good and these approximations turned out
to be not very crucial. We believe that an incorporation of
forward-backward asymmetries into the fits may be worthwhile and to this end
we encourage experimenters to systematically present raw data in addition
to extracted or bare quantities. There is another benefit to it:
the extraction of bare quantities often requires at some point assumptions
such as the validity of the Standard Model when data from outside are input.
This usually leads to small
effects, but it destroys the consistency of global fits to high precision data.

\section*{Acknowledgements}
It is a pleasure to thank Paul Langacker for his suggestion to study the
topic presented in this article, for his careful reading of the manuscript and
for numerous very helpful suggestions and comments.
This work was supported by the Texas National Laboratory Research Commission,
by the D.O.E. under contract DE-AC02-76-ERO-3071 and
by the Deutsche Forschungsgemeinschaft.

\newpage
\appendix
\section{Approximations for lineshape fits}
\label{approx}
In this appendix we list the simplifications we used for our fits:
\begin{enumerate}
\item
   We worked in the improved Born approximation, i.e., box contributions were
   neglected and the $s$-dependent effective couplings were assigned their
   values at $\sqrt{s} = M_Z$.
\item
   In the $Z$ interference terms we kept the effective mixing angles fixed. The
   axial-vector couplings are absorbed into the widths
   (see appendix~\ref{ISR}).
\item
   For hadronic cross sections we neglected systematic uncertainties from
   selection cuts, efficiencies and backgrounds, which are small compared
   to the luminosity error.
\item
   In order to avoid the use of a high dimension correlation matrix we
   proceeded in the following way: We defined a luminosity scale factor by
\be
    L = \frac{{1\over (\Delta L)^2} +
        \sum {\sigma_{\rm had}^{\rm fit} \sigma_{\rm had}^{\rm exp} \over
             (\Delta {\sigma_{\rm had}^{\rm exp}})^2}}
             {{1\over (\Delta L)^2} +
        \sum {{\sigma_{\rm had}^{\rm fit}}^2 \over
             (\Delta {\sigma_{\rm had}^{\rm exp}})^2}},
\ee
   where the sum is over the scan points. We then substituted for hadrons
   {\em and\/} leptons
   $\sigma^{\rm fit} \rightarrow L \sigma^{\rm fit}$
   in our $\chi^2$ function and added the term
   \be
     {(L-1)^2 \over (\Delta L)^2},
   \ee
   where the luminosity error $\Delta L = 0.006$ is taken to be fully
   correlated between the different years. For $\Delta L \rightarrow \infty$,
   i.e., disregarding the constraint from the direct luminosity measurement,
   this procedure corresponds to a determination of the luminosity from the
   cross section data.
\item
   We chose a similar treatment for the systematic uncertainties (excluding
   the luminosity error) for $\mu$ and $\tau$ final states, which we assumed
   to be mutually uncorrelated. Again, these uncertainties were taken to be
   completely correlated between different years and equal, although the
   1990 errors were slightly higher than the ones from 1991 and 1992.
\item
   Initial state radiation is included in one loop exponentiated form
   except for the pure QED contribution where the exponentiation can be
   omitted~\cite{yellowrep}. We present the explicit formulae including
   four-Fermi terms and their interferences in appendix~\ref{ISR}.
\item
   Some convolution integrals must be
   extended to a larger region if analytical formulae are desired~\cite{Cahn}.
\item
   We did not include $e^+$ $e^- \rightarrow e^+$ $e^-$ data to avoid
   complications due to $t$-channel exchange.
\item
   We did not include any forward-backward asymmetry data. Their inclusion
   is straightforward but somewhat tedious, since another class of convolution
   integrals has to be calculated.
\item
   We did not include total cross section data of 1993, since they are not yet
   available in published form.
\end{enumerate}
It turned out a posteriori that these simplifications have little impact
on the fit results. If we compare our Standard Model fit with the one by the L3
collaboration\footnote{The quoted L3 result of $\Gamma_l$ is the weighted
average
of $\Gamma_\mu$ and $\Gamma_\tau$, whereas the $\chi^2$ value
includes $e^+$ $e^-$ final states, as well. Our quoted errors are those
returned by the minimization routine MINUIT.}~\cite{L3},
\be
\ba{rccc}
           &&{\rm our\,\,\, fit} & {\rm L3}  \\

             M_Z\, [{\rm GeV}] &=&
         91.193 \pm 0.006 &\;\;\; 91.195 \pm 0.006 \pm 0.007\; ({\rm LEP}), \\

        \Gamma_Z\, [{\rm GeV}] &=&
       2.496 \pm 0.010 &\;\;\;\;\;  2.494 \pm 0.009 \pm 0.005\; ({\rm LEP}) ,\\

         \sigma_0\, [{\rm nb}] &=& 41.42  \pm 0.20\,\,  & 41.41  \pm 0.26, \\

        \Gamma_l\, [{\rm MeV}] &=& 83.62  \pm 0.35\,\, & 83.55  \pm 0.60, \\

	\chi^2 &=& 41/44 & 53/60,

\ea
\ee
we see that the central values agree within better than 0.1\%.
We reached this precision without any numerical integration or matrix
manipulation so that the fit runs are very fast.
The precision can be further improved, if desired.

\section{Initial state radiation}
\label{ISR}

We treated initial state radiation using the approximations described in
reference~\cite{yellowrep}. Before convolution, the total cross section
formula reads
\be
\ba{c}
   \sigma_f^\prime(s) =
               (\frac{s^2 C_{I4F} + s (C_R + C_I - M_Z^2 C_{I4F}) - M_Z^2 C_I}
                    {(s - \tilde{M}_Z^2)^2 + \tilde{M}_Z^2 \tilde{\Gamma}_Z^2}
             + {C_Q \over s} + s C_{4F} + C_{IQ4F}) \\
               (1 + {3\over 4 \pi} \alpha (M_Z) Q_f^2)(1 + \delta_{QCD})
               \sqrt{1-4 {m_f^2/M_Z^2}},
\ea
\ee
where for instance for a pure vector type four-Fermi contribution
\be
\ba{rcl}
   C_R &=& \frac{12 \pi \Gamma_e \Gamma_f}{M_Z^2 + \Gamma_Z^2}, \\
   C_Q &=& {4\over 3} \pi Q_f^2 \alpha (M_Z)^2 N_c, \\
   C_{4F} &=& {4 \pi N_c V_f^2 \over 3 \Lambda^4}, \\
   C_I &=& \frac{8 \pi \alpha(M_Z) Q_e Q_f M_Z \sqrt{N_c \Gamma_e \Gamma_f}}
              {M_Z^2 + \Gamma_Z^2}
         \frac{g_V^e}{\sqrt{{g_V^e}^2+1}}\frac{g_V^f}{\sqrt{{g_V^f}^2 + 1}}, \\
   C_{I4F} &=& \frac{8 \pi V_f M_Z \sqrt{N_c \Gamma_e \Gamma_f}}
                  {\Lambda^2 (M_Z^2 + \Gamma_Z^2)}
         \frac{g_V^e}{\sqrt{{g_V^e}^2 + 1}}\frac{g_V^f}{\sqrt{{g_V^f}^2+1}}, \\
   C_{I4FQ} &=& {8 \pi Q_e Q_f \alpha (M_Z) V_f N_c \over 3 \Lambda^2}, \\
   \tilde{M}_Z &=& \frac{M_Z}{\sqrt{1 + \Gamma_Z^2/M_Z^2}}, \\
   \tilde{\Gamma}_Z &=& \frac{\Gamma_Z}{\sqrt{1 + \Gamma_Z^2/M_Z^2}}.
\ea
\ee
The partial widths are with QED, QCD and mass corrections removed;
$g_V^f$ denotes the vector coupling of fermion $f$; and $\delta_{QCD}$
is the QCD correction, which in case of $b$-quarks is $m_b$ and
$m_t$-dependent~\cite{QCDT2} and a weighted average of the corrections
to the vector and axial-vector partial widths has to be used.

Initial state radiation is included by computing the convolution integral
\be
   \sigma_f (s) =
   \int\limits_{0}^{1-{s_0\over s}} {\rm d} x \sigma_f^\prime[s(1-x)] G(x),
\ee
where $s_0$ is taken to be $4 m_f^2$ for leptons and $(10\; {\rm GeV})^2$ for
hadrons. $G(x)$ is the radiator function;
to one-loop exponentiated approximation it is given by~\cite{BBN}
\be
   G(x) = \beta x^{\beta -1} \delta^{V+S} - \frac{\alpha}{\pi}(2-x)(L-1),
\ee
where
\be
\ba{rcl}
   L &=& {\rm ln} {s\over m_e^2}, \\
   \beta &=& {2 \alpha\over \pi} (L-1), \\
   \delta^{V+S} &=& 1 + {\alpha\over \pi} ({3\over 2} L + {\pi^2\over 3} -2).
\ea
\ee
The final result is
\be
\ba{c}
   \sigma_f(s)=\{(C_{I4F}+{C_R+C_I-M_Z^2 C_{I4F}\over s}-{M_Z^2 C_I\over s^2})
              (J_\beta \delta^{V+S} - {\beta \over 2} J_1) \\
            - (2 C_{I4F} + {C_R + C_I - M_Z^2 C_{I4F} \over s})
              ({\beta \over \beta + 1} J_{\beta + 1} \delta^{V+S}
            -  {\beta \over 2} J_2) \\
            + C_{I4F} ({\beta \over \beta + 2} J_{\beta + 2} \delta^{V+S}
            - {\beta \over 2} J_3) \\
            + {C_Q\over s} [ 1 + {\alpha\over \pi} L \left( {1\over 2} +
              {s_0\over s}-
              {\rm ln}{s_0\over s} + 2\; {\rm ln}(1 - {s_0\over s})\right)
            + {\alpha\over \pi} \left( {\pi^2\over 3} - 1 - {s_0\over s} +
              {\rm ln}{s_0\over s}-2\;{\rm ln}(1-{s_0\over s}) \right) ] \\
            + s C_{4F} [\frac{\beta {s_0\over s} + 1}{\beta + 1}
              (1 - {s_0\over s})^\beta \delta^{V+S} - {\beta\over 2}
              ({5\over 6} - {s_0^2\over 2s^2} - {s_0^3\over 3s^3})] \\
            + C_{I4FQ} [(1-{s_0 \over s})^\beta \delta^{V+S} - {\beta\over 2}
              ({3\over 2} - {s_0\over s} - {s_0^2\over 2 s^2})]\} \\
            (1 + {3\over 4 \pi} \alpha (M_Z) Q_f^2)(1 + \delta_{QCD})
            \sqrt{1-4 {m_f^2/M_Z^2}}.
\ea
\ee
In order to compute the $J_n$ we extend the integration region
to $s_0 = 0$~\cite{yellowrep},
\be
  J_n = \int\limits_0^1 {\rm d}x \frac{x^{n-1} (2-x)}{(x+a)^2 + b^2},
\ee
and we use the abbreviations
\be
\ba{l}
  a = {\tilde{M}_Z^2 \over s} - 1, \\
  b = {\tilde{M}\tilde{\Gamma}_Z \over s}.
\ea
\ee
The result is
\be
\ba{l}
  J_1 = {2 + a \over b} A - {B\over 2}, \\
  J_2 = {a^2 + 2 a - b^2 \over b} A + (1 + a) B - 1, \\
  J_3 = {2a^2+a^3-2b^2-3ab^2 \over b}A+{b^2-3a^2-4a \over 2}B+2a+{3\over 2},
\ea
\ee
with
\be
\ba{l}
   A = \arctan {a+1 \over b} - \arctan {a \over b}, \\
   B = {\rm ln} \frac{a^2 + b^2 + 2 a + 1}{a^2 + b^2}.
\ea
\ee
The integration region for $J_\beta$ with $\beta < 2$ is
extended even further~\cite{yellowrep,Cahn},
\be
   J_\beta = \beta \int\limits_0^\infty {\rm d}x
             \frac{x^{\beta - 1}}{x^2 - 2 \eta x \cos\zeta + \eta^2}
           = \eta^{\beta - 2} \frac{\pi \beta \sin[(1 - \beta) \zeta]}
                                   {\sin \pi \beta \sin \zeta},
\ee
and we defined
\be
\ba{l}
  \eta = \sqrt{a^2 + b^2}, \\
  \cos\zeta = {a \over \eta}.
\ea
\ee
We checked that the extension of the integration domain can be understood
as an expansion in $\eta$ (which for LEP energies is of order
${\cal O} (\Gamma_Z/M_Z)$) and keeping only negative powers. $J_{\beta+2}$
has no negative powers of $\eta$ and we have to keep the $\eta$ independent
term to insure that it vanishes in the limit
$\beta \rightarrow 0$. We obtain ($0 \leq \beta < 1$),
\be
  J_{\beta + 2} = {\beta + 2 \over \beta} [1 - \eta^2 J_\beta -
                  2 \cos \zeta {\beta \over \beta + 1} \eta J_{\beta + 1}].
\ee
Actually, in case of $J_\beta$ above, one should for the sake of
self-consistency keep at least the $\eta$-independent term as well, since it is
comparable to the $J_n$ terms. We find up to terms linear in $\eta$
($\beta < 3$),
\be
  J_\beta = \beta\left[\eta^{\beta - 2} \frac{\pi \sin[(1 - \beta) \zeta]}
            {\sin \pi \beta \sin \zeta} + {1 \over \beta - 2} -
            {2 \over \beta - 3} \eta \cos\zeta + {\cal O} (\eta^2)\right].
\ee

%\vspace{4.0ex}

%\begin{thebibliography}{99}

\onecolumn

\end{document}